# A Selection Criterion for Patterns in Reaction-Diffusion Systems

Tatiana T. Marquez-Lago[1,2,*] and Pablo Padilla[3]

[1] STMC and Department of Mathematics, University of New Mexico, Albuquerque, NM, 87131, USA.

[2] Present address: Integrative Systems Biology Unit, Okinawa Institute of Science and Technology, 1919-1 Tancha, Onna-son, Kunigami, Okinawa 904-0412, Japan.

[3] IIMAS, Universidad Nacional Autónoma de México, Circuito Escolar. Cd. Universitaria, 04510, México D.F., Mexico.

[*] To whom correspondence should be addressed: tatiana.marquezlago@gmail.com

## ABSTRACT

Alan Turing's work in Morphogenesis has received wide attention during the past 60 years. The central idea behind his theory is that two chemically interacting diffusible substances are able to generate stable spatial patterns, provided certain conditions are met. Turing's proposal has already been confirmed as a pattern formation mechanism in several chemical and biological systems and, due to their wide applicability, there is a great deal of interest in deciphering how to generate specific patterns under controlled conditions. However, techniques allowing one to predict what kind of spatial structure will emerge from Turing systems, as well as generalized reaction-diffusion systems, remain unknown. Here, we consider a generalized reaction diffusion system on a planar domain and provide an analytic criterion to determine whether spots or stripes will be formed. It is motivated by the existence of an associated energy function that allows bringing in the intuition provided by phase transitions phenomena. This criterion is proved rigorously in some situations, generalizing well known results for the scalar equation where the pattern selection process can be understood in terms of a potential. In more complex settings it is investigated numerically. Our criterion can be applied to efficiently design Biotechnology and Developmental Biology experiments, or simplify the analysis of hypothesized morphogenetic models.

# 1. INTRODUCTION

Turing models and general reaction-diffusion systems have been used to study mechanisms leading to emergent spatial patterns. Such studies have proved useful in a wide range of fields, including Biology, Chemistry, Physics, Ecology and Economics (see [1] and references therein). Moreover, their applicability to biological and chemical pattern formation processes has been firmly established since Alan Turing first proposed it as the basis of morphogenesis [2, 3].

Patterns arising in reaction-diffusion processes can be observed in well-known oscillatory models such as the Brusselator [4, 5] and the Oregonator model [6] of the Belusov-Zhabotinsky chemical reaction. One can also observe distinct geometric patterns through the Schnakenberg and Brandeisator models, of the CIMA reaction [1], among many others. In Biology, Turing models have been proposed to describe developmental processes such as skin pigmentation patterning [7, 8], hair follicle patterning [9], and skeletal development in limbs [10]. Remarkably, synthetic multicellular systems have been programmed *de novo* to generate simplified patterns using quorum sensing mechanisms [11, 12], envisioning future applications in tissue engineering [13]. More recently, tomography studies of microemulsions have revealed three-dimensional Turing patterns [14]. Thus, due to their large applicability in several research fields, understanding the relationship between reaction-diffusion parameters and specific patterns becomes essential.

In his seminal work, Turing considered two distinct diffusible substances interacting with each other The key idea behind Turing models is the so-called diffusion-driven instability, for which one needs to first consider an asymptotically stable equilibrium without diffusion. Then, if diffusion is incorporated, the equilibrium might become unstable if spatial perturbations are considered, undergoing a symmetry breaking process. As a consequence, spatially nonhomogeneous stable patterns can be produced. In fact, it is the ability of Turing patterns to regenerate autonomously that gives them great utility in applications such as tissue engineering and developmental processes [13, 15].

Turing patterns might be structurally complicated but, roughly speaking, exhibit two kinds of generalized structures: spots or stripes. So far, heuristic criteria have been proposed, with restrictions on reactive terms (e.g. [16, 17]). The main purpose of this paper is to propose an analytic selection criterion aimed at predicting patterns for general reaction-diffusion systems, depending on the nonlinearities involved in the reaction terms. We will illustrate these ideas with two different types of pattern-generating reaction-diffusion systems, namely, the Gierer-Meinhardt activator-inhibitor model (a short-range positive feedback coupled to a long-range negative feedback) [18] and the Fitzhugh-Nagumo equations [19]. For such, we shall consider the general form

$$\frac{\partial u}{\partial t} = D_1 \nabla^2 u + f(u,v)$$

$$\frac{\partial v}{\partial t} = D_2 \nabla^2 v + g(u,v)$$

(1)

in a two dimensional spatial domain $\Omega$, supplemented with zero flux boundary conditions. This model has been extensively used in many contexts and is by now standard (see for instance [2]).

Generally, when considering a Gierer-Meinhardt system, the observed patterns are either spots or stripes. Correspondingly, one can observe either spots or labyrinthic-like patterns in Fitzhugh-Nagumo systems. In fact, in [20], a systematic analysis for a generic cubic nonlinearity is carried out for the first type of systems, and it is shown that the formation of spots and stripes depends on the presence or absence of a quadratic term respectively (see also [21] for more recent works on the subject). For the simplest case, namely a scalar equation in a one dimensional domain, a full analytical solution can be given in terms of the qualitative structure of the nonlinearity. Although there are no spots or stripes in this case, a simple interpretation can be given. Looking for steady state solutions leads to a classical problem in mechanics and an energy function is well defined.

At this point, a few words regarding both terminology and previous results are necessary. Throughout this paper, we will refer to the scalar case when we consider one single equation (that is only one dependent variable or morphogen, say $u$, is present). In this case, when the domain $\Omega$ is convex, Casten and Holland proved that the only stable patterns are spatially constant ([22] and in [23] a more detailed account of these results is given). Notice that this necessarily means that all nontrivial patterns (i.e. nonhomogenous patterns for the scalar case) are unstable and therefore difficult to find numerically. This will be important in the numerical examples presented later on. Additionally, we will refer to the physical space variables (i.e. those appearing in the reaction-diffusion system implicitly in the Laplace operator) as physical or independent variables. Finally, when we talk about the morphogens or dependent variables, we sometimes will use the terminology of state variables.

With this terminology in mind, let us go back to the definition of an energy function and its use as a selection criterion. Roughly speaking, the criterion for the scalar case states that if the associated energy of the system is bistable (i.e. two minima) but has a unique global minimum, then spots will be formed, corresponding to the existence of homoclinic solutions (again in analogy with the mechanical case). Indeed, by solving the equation for special nonlinearities explicitly, it becomes clear that a homoclinic solution represents a spot, and that a heteroclinic solution (a front) represents a stripe. This has a simple interpretation in the language of phase transitions. Consider a two phase system with an energy function having two minima (corresponding precisely to the stable phases), a so called double-well potential. If the minima are not at the same height, it means that one of the phases, say A, has less energy than B. If the two phases coexist, it will be energetically speaking more efficient to have a 'background' of A surrounding patches of B. If one considers a usual term in the energy that penalizes the region of transition between phases, it is natural, as it actually occurs, that the patches will be circular (namely spots). On the other hand, if the roles of A and B are inverted, then the situation is completely analogous, but in this case spots of phase A would appear, surrounded by a background of B. However, in the case when the minima heights are identical, there is no preferred phase in terms of the energy and the formation of more complicated structures, for instance, stripes. The scalar case is atypical in the sense that the results by Casten and Holland mentioned above establish that no spatially nonhomogeneous patterns can be formed. In this respect, the formation of spots or stripes can only be understood in the metastable sense since, if a stable pattern is eventually achieved, it will be constant (see the numerical simulations presented later on).

Our proposed criterion is a natural generalization of this fact. Using phase plane analysis and standard methods in differential equations and mechanics, we show that an analog of the energy of the system provides a clear way of predicting the pattern that will be formed. In two spatial dimensions and for general reaction diffusion systems, an extension of this criterion can be obtained by reinterpreting the results in one dimension in a probabilistic way, proposing a function that plays the role of energy. If the system itself has variational structure, then the same reasoning can be made, since a natural energy function exists. The important fact is that, even in the case where there is no variational structure (and therefore, no a priori energy function given), it can be constructed. It will be given as the solution of a Fokker-Planck (FP) equation associated with the original reaction-diffusion system.

In dynamical terms and for the scalar case, spots will be formed if the energy has a unique global minimum, corresponding to the existence of a homoclinic solution (a typical spot). Indeed, by solving the equation explicitly for special nonlinearities it becomes clear that a homoclinic solution represents a spot, and that a heteroclinic solution (a transition layer) represents a stripe. When these special cases are considered on the plane, they correspond to radial solutions for the energy with a single global minimum, and to transition layers for the double-well energy, with the bottoms of the two wells having the same height.

That the actual coherent structures that are formed correspond to spots or stripes is due to the fact that besides the potential energy term, there is a gradient term that penalizes the transitions themselves. In other words, minima are achieved not only when the potential energy is made small, but also when the transition region between phases is kept to a minimum. This is analogous to an isoperimetric problem of minimizing the perimeter of a domain of a given area. For a precise mathematical formulation of these statements we refer to [24].

In order to construct the analogue of the energy in the general case, we write down an associated Fokker-Planck equation for the transition probability, $\rho$, of the associated dynamical system, interpreting diffusion terms in the standard probabilistic way. Then we propose that -$\rho$ plays the role of the energy function, and the fact that spots or stripes are formed depends on whether it has one or more than one global minima. We will examine these criteria by means of numerical simulations. From the analytic point of view we show that, under suitable assumptions, -$\rho$ is a Lyapunov function for the reaction-diffusion system, in the same way the energy is a Lyapunov function for the one dimensional case or, in general, when the system has variational structure. This provides a rigorous proof to the intuitive discussion on phase transitions above. The proofs are based on a well-known fact about invariant regions for reaction-diffussion systems (see below and [25] for more details). It is worth mentioning reaction-diffusion systems do not in general possess a Lyapunov function, as evidenced by the fact that they may show periodic or chaotic time dependence. However, in what follows, we do not imply global Lyapunov functions.

We point out that our proposal generalizes the work presented in [20], in which reaction-diffusion models containing quadratic and cubic reaction terms were studied. In fact, reinterpreting these results with our criterion is straightforward, since the absence of an even term in the equation implies the absence of an odd term in the corresponding potential and therefore, the existence of two global minima, yielding stripes formation. Nonetheless, as soon as an odd term is present, there is only one global minimum, yielding spots. It is also important to mention here that these facts suggest a general result that is observed in our simulations. In these simulations, since the presence of a cubic term is generic, spot formation is expected to be the robust option. This is also true for our criterion, since the generic situation is that -$\rho$ has only one global minimum.

Lastly, for the Fitzhugh-Nagumo system we will carry out analogous simulations. The structures that are formed are spots or labyrinth-like structures and our numerical simulations suggest that our criterion is still valid, showing versatility across different types of Turing systems.

## 2. RESULTS

### 2.1. Preliminaries for selection criterion: analytic solution for the scalar case in 1D

Our approach can be easily understood if we start by analyzing the limiting case of the full reaction-diffusion system. Namely, when the diffusion of $v$ is much faster in comparison to that of $u$ (so that $v$ can be considered as essentially constant, see [23] and references therein). In fact, in some cases, It is well known that the dynamics of morphogens are essentially determined by only one of them [23]. However, it should be stressed that stability is a different issue (for instance, there are no Turing patterns for the scalar case - see below for a more detailed discussion on this and the observations in the previous section). If it were the case that species $u$ essentially determines the dynamics of both morphogens, system (1) could be reduced to

$$\frac{\partial u}{\partial t} = D \nabla^2 u - f(u), \tag{2}$$

the so-called shadow equation.

In this case, when the spatial domain is an interval and stationary patterns are looked for, the equation can be explicitly solved. In fact, the equation is equivalent to Newton's equation for the motion of a particle under the action of a potential. This is the case, specifically, since we are looking for stationary patterns, and then $\frac{\partial u}{\partial t} = 0$. Moreover, in one spatial dimension, the shadow equation would further reduce to $\frac{d^2 u}{dx^2} = f(u)$.

So the role of the potential is played by $F(u) = -\int f(u) du$, and the spatial variable $x$ plays the role of time. This equation can be explicitly solved if we multiply it by $du/dx$, as it is standard in classical mechanics. Since $F'(u) = -f(u)$,

$$\frac{d}{dx}\left(\frac{1}{2}\left(\frac{du}{dx}\right)^2 + F(u)\right) = 0. \tag{3}$$

This implies that the quantity in parentheses is constant, i.e.

$$\frac{1}{2}\left(\frac{du}{dx}\right)^2 + F(u) = E. \tag{4}$$

As a matter of fact, in the mechanical analog, $E$ corresponds to the energy of the system. Solving now for $du/dx$ and integrating, we have the solution given implicitly as

$$\int_{u(a)}^{u(x)} \frac{du}{\sqrt{2(E - F(u))}} = x - a, \tag{5}$$

where we take the interval on the $x$ axis to be $[a, b]$.

So far, it is straightforward to do these derivations. So, let us now focus on the case when $f$ has two stable equilibria and $uf(u) > 0$ for large values of $u$. This is the first nontrivial case to study since, for a single well, solutions would be trivial and would correspond to the (unique) minimum of a convex functional. One simple yet illustrative example is $F(u) = (1 - u^2)^2 + bu$, where $-f(u) = -4u(1 - u^2) + b$, and $b$ varies. The case $b = 0$ corresponds to a bistable (double well) potential which goes to plus infinity with $u$. We can also analyze the cases where $b \neq 0$. The formation of spots/ inverted spots or stripes can be deduced from the phase plane and correspond to homoclinic or heteroclinic solutions respectively (FIG. 1). Alternatively, and recalling our discussion about phase separation, patterns would correspond to cases where there is a preferred phase with minimal energy.

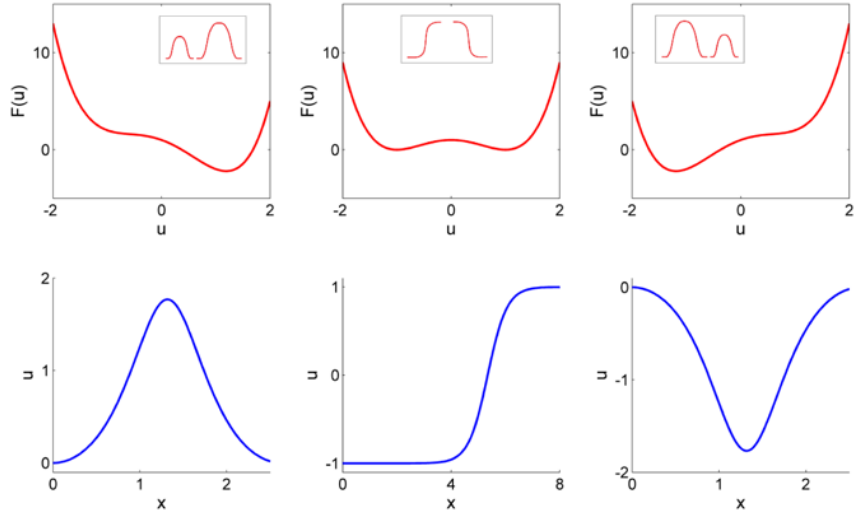

FIG. 1. Solutions of the stationary shadow equation in one spatial dimension, $u''(x) + 4u(x)[1 - u(x)^2] - b = 0$. Upper panels show associated potentials, while lower panels show stationary solutions of the shadow equation. Column cases correspond to (left) b = -2, (middle) b = 0, (right) b = 2. Inset figures of upper panels depict qualitative homoclinics (left and right panels) and symmetric transitions (center panel).

The last assertion might seem strange, since it makes no sense to talk about spots or stripes in systems embedded in one spatial dimension. However, in two spatial dimensions, it can be shown that nontrivial minimal solutions are radially symmetric. For the case of homoclinics, they would correspond to rotations of the homoclinic solution in one dimension (this follows from well-known results about radial symmetry by Gidas, Ni and Nirenberg [26]). Correspondingly, for the heteroclinic case, minimal energy nontrivial solutions depend on one variable only and therefore exhibit a stripe-shaped structure. This assertion and similar mathematical issues have been studied in detail in the context of phase transition models and regularizations of the minimal surface problem (see for instance [24, 27] and references therein). We illustrate this in FIG. 1, where homoclinics correspond to spot-like solutions. The smaller homoclinic is associated with the well of least depth, and the larger one with the deeper well, respectively. However, it is worth noting that, since the lower amplitude homoclinics correspond to larger energies, they are also less stable. Also in FIG. 1 (middle, inset) there are two symmetric transitions, corresponding to stripes.

In the case of two or three spatial dimensions the corresponding equation cannot be explicitly solved in general, but its solutions can still be characterized as the critical points of a functional. The latter would be the Euler-Lagrange equation of the "action"

$$\int_\Omega |\nabla u|^2 + F(u) \, dx, \tag{6}$$

for which a qualitative analysis similar to the one-dimensional case can be carried out, as mentioned in the preceding paragraph. For additional information on the topic, we refer the reader to the vast literature on variational methods and formation of localized structures (see [28]). For readers not familiar with the results and techniques of the calculus of variations, we refer to the Methods section (cf. Calculus of Variations).

So, if one considers system (1), the approach described above in one dimension can be extended to account for special nonlinearities corresponding to the gradient of a function, i.e. that have a variational structure. However, a full analytic treatment may not be always possible and alternative (numerical) methods should be adopted. Nevertheless, the interested reader should refer to the variational case section for complete details.

## 2.2 Selection criterion in a nutshell

In this paper we show that a pattern selection criterion can be obtained even when the system does not have a gradient structure. The idea is that the role of the potential is now played by minus the transition probability associated with the system

$$\frac{du}{dt} = f(u,v)$$
$$\frac{dv}{dt} = g(u,v)$$
(7)

when subjected to a random perturbation determined by the diffusion coefficients. That is, the stochastic differential equations

$$du = f(u,v)dt + \sqrt{2D_1}\,dW_1$$
$$dv = g(u,v)dt + \sqrt{2D_2}\,dW_2$$
(8)

where we have written the vector field $V(u,v) = (v_1(u,v), v_2(u,v)) = (f(u,v), g(u,v))$, and $W_1, W_2$ are the components of two dimensional standard Brownian motion.

Then, the transition probability density $\rho(u,v,t)$, i.e. the probability of finding the system in the state $(u,v)$ at time $t$, satisfies the Fokker-Planck equation

$$\frac{\partial \rho}{\partial t} + \nabla \cdot (\rho V) = \frac{\partial \rho}{\partial t} + \frac{\partial}{\partial u}[f(u,v)\rho(t,u,v)] + \frac{\partial}{\partial v}[g(u,v)\rho(t,u,v)] = D_1 \frac{\partial^2 \rho}{\partial u^2} + D_2 \frac{\partial^2 \rho}{\partial v^2} \quad (9)$$

with zero flux boundary conditions (cf. Reference [29] for details on the relationship of (1), (7)-(9)). It is worth noting (9) is a special case of the Feynman-Kac formula, for which a solution can be analytically found. As it was pointed out before, we claim that $\rho$ plays the role of the potential and it allows us to predict the appearance of a particular pattern in a given system.

It is worth noting that, we are implicitly assuming that the intensities of noise terms in the state space entering the stochastic differential equations (and leading to the diffusion coefficients in the FP equation (9)) are related to the diffusion coefficients of the original reaction-diffusion equation (1). Even if this is not easy to justify, it is not an unreasonable assumption if one considers that the Fokker-Planck equation describes a "diffusion" process in an abstract state space and that the reaction-diffusion process occurs in physical space. Thus, locally, in the physical space, one may think that at every point an abstract diffusion process in the state variables is taking place. After a suitable choice of local spatial coordinates one may locally take such coordinates as state variables in a minimal coupling way.

We now illustrate that, depending on a relevant parameter in the activator-inhibitor system which might be negative, positive or zero, the stationary transition probability will have one or two global maxima. These cases correspond to the formation of spots, inverted spots or stripes. This is also concluded by numerically solving the corresponding Turing system. Similarly, we show that for a proposed Fitzhugh-Nagumo system, depending on such a parameter, which might be positive or negative, the stationary transition probability has one or two maxima, corresponding to the formation of spots.

## 2.3 Stripes versus spots revisited

We start by considering the scalar equation

$$\frac{\partial u}{\partial t} = D \nabla^2 u + 4u(1 - u^2) - bu^2 \tag{10}$$

and present numerical evidence illustrating our selection criterion, followed up by its analytic proof. In the Preliminaries section (section 4.1), a similar case was treated analytically, albeit in one spatial dimension. We consider important to show the numerical results in the two dimensional case (i.e. the lattice, in two spatial dimensions) out of completeness. As was discussed before, nontrivial minimal solutions are found to be radially symmetric for homoclinics, while for heteroclinics minimal energy nontrivial solutions exhibit a stripe-shaped structure.

In this case, we solved the associated Fokker-Planck equation

$$\frac{d\rho}{dt} + \frac{\partial}{\partial u} \left[ \left( 4u(1 - u^2) - bu^2 \right) \rho \right] = 0.1 \frac{\partial^2 \rho}{\partial u^2} \tag{11}$$

and show that, depending on parameter $b$, the stationary transition probability yields one or two global maxima, corresponding to the formation of spots, inverted spots or stripes. These results are shown in FIG. 2.

Then, in order to check whether the parameters correspond to the expected patterns, we discretized our scalar equation (10) and solved it using a rectangular grid ($14 \times 14$ units) with $10^4$ grid points. We set a diffusion coefficient of $D = 0.1$ and adopted random initial conditions. For additional details on the numerical method, please refer to the Methods section (cf. Numerical discretization of reaction-diffusion system). We observed that, for parameters that determine spots and inverted spots (e.g. $b = -0.02$ and $b = 0.02$, respectively), the solution becomes constant as time progresses (i.e. a trivial spot or inverted spot). This is due to the fact that in one dimension nontrivial solutions are unstable. Sample simulations are shown in FIG. 3. On the other hand, when we use the parameters that determines stripes, $b = 0$, the stationary solution can be a trivial spot or inverted spot, and the stationary solution is solely determined by initial conditions.

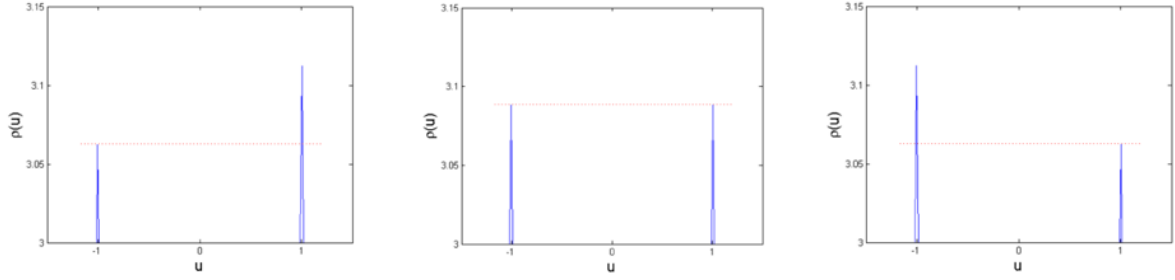

FIG. 2. Magnification of associated Fokker-Planck solutions of 1-D Turing model (10), corresponding to one morphogen in two spatial dimensions. Original solutions were obtained over the domain $x \in [-3.15, 3.15]$. Cases correspond to parameters (left) b = -0.02, (middle) b = 0, and (right) b = 0.02. The Fokker-Planck equation was solved in Comsol, until steady state (total time T = 1000) with a time step of 0.01, and zero-flux boundary conditions. The initial condition was defined as $u(x, t = 0) = e^{-x^2}$ over the domain, and a diffusion coefficient $D = 0.1$ was used. Dotted lines are used to emphasize differences between local optima.

Moreover, depending on initial conditions, a transition layer can also be formed. For parameters that determine spots and inverted spots the latter is in fact a moving front, yielding at steady state the aforementioned trivial solution. On the other hand, when using parameters that determine stripes, the transition layer constitutes a stationary solution (the time evolution only serves to align it to be normal to the boundary). A sample simulation is shown in FIG. 3.

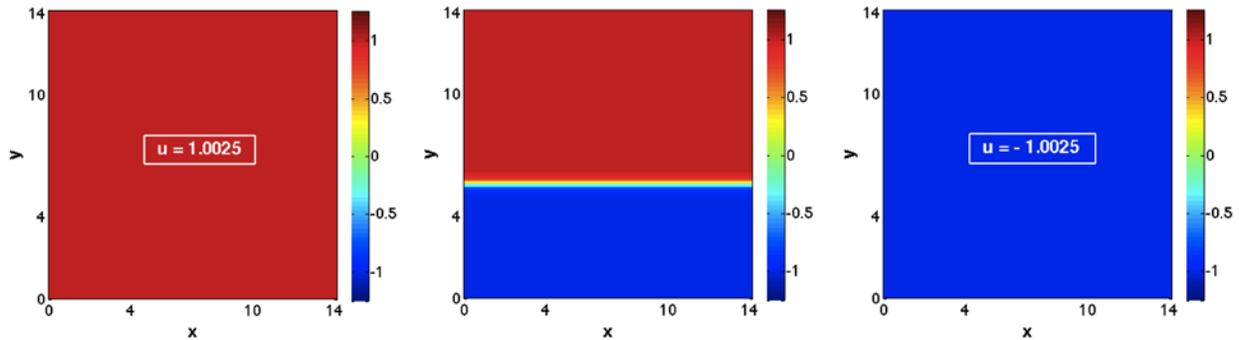

FIG. 3. Sample stationary solution of 1-D Turing model (10) in two spatial dimensions. Panels correspond to (left) inverted spots with b = -0.02, (center) stripes with b = 0, and (right) spots with b = 0.02. All other model/simulation parameters are indicated in the text. Simulations were run until a numerical steady state was reached (maximum difference in any lattice point less than 1e-10).

Now, in the scalar equation scenario, if we set $V = \nabla(-\varphi)$ and consider the whole range of $\mathbb{R}$, the stationary solution of the Fokker-Planck equation

$$\frac{\partial \rho}{\partial t} + \nabla \cdot (\rho V) = D \frac{\partial^2 \rho}{\partial u^2} \tag{12}$$

is given explicitly by

$$\rho(u) = \exp\left(-\frac{\varphi(u)}{D}\right) \tag{13}$$

Since $-\rho$ and $\varphi$ have the same critical points' structure, we have shown our criterion coincides with the solution obtained in the mechanical analog derived in section 4.1. In other words, and qualitatively speaking, one can equally consider $\rho$ or $\varphi$ in terms of the number of global minima.

We will now turn to a classical Turing system of two equations

$$\frac{\partial u}{\partial t} = D \nabla^2 u + 0.6\, u - r\, v + q u^2 - u^3$$

$$\frac{\partial v}{\partial t} = \nabla^2 v + 1.5\, u - 2v, \tag{14}$$

in two spatial dimensions, where we again solved the associated Fokker-Planck system of equations. We show that, depending on parameter $q$, the stationary transition probability yields one or two global maxima (FIG. 4), corresponding to the formation of spots, inverted spots or stripes. For this two morphogen case we discretized system (14) using a diffusion coefficient of $D = 0.04$, random initial conditions, and a rectangular grid (20 × 20 units) with $10^4$ grid points (see Methods section for further details). Once again, for parameters that determine spots and inverted spots (e.g. $q = -2$ and $q = 2$ respectively) one can see that as time progresses the solution becomes constant (i.e. a trivial spot). Sample simulations are shown in FIG. 5, where it should be noted the value of parameter r was also varied, to show the desired patterns without changing the size of the simulation domain. However, if one uses $q = 0$, the obtained stationary pattern is stripes. The geometry of the stripes need not be straight, and stripes detach and bind to each other until all borders are normal to the boundary, soon after which a steady state is reached. Sample simulations are shown in FIG. 5.

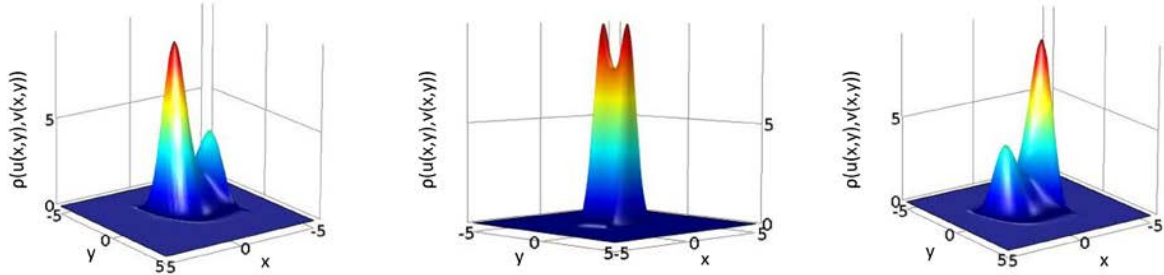

FIG. 4. Associated Fokker-Planck solutions of 2-D Turing model (14), corresponding to two morphogens in two spatial dimensions. Cases correspond to (left) q = -2 and r = 4.5, (middle) q = 0 and r = 1, (right) q = 2 and r = 4.5. The Fokker-Planck equation was solved in Comsol, until steady state (total time T = 1000) with a time step of 0.01, and zero-flux boundary conditions. The initial condition was defined as $u(x, y, t = 0) = e^{-(x^2+y^2)}$ over the domain $x, y \in [-5,5]$, and a diffusion coefficient $D = 0.04$ was used for morphogen $u$ (morphogen $v$ has a diffusion coefficient equal to 1).

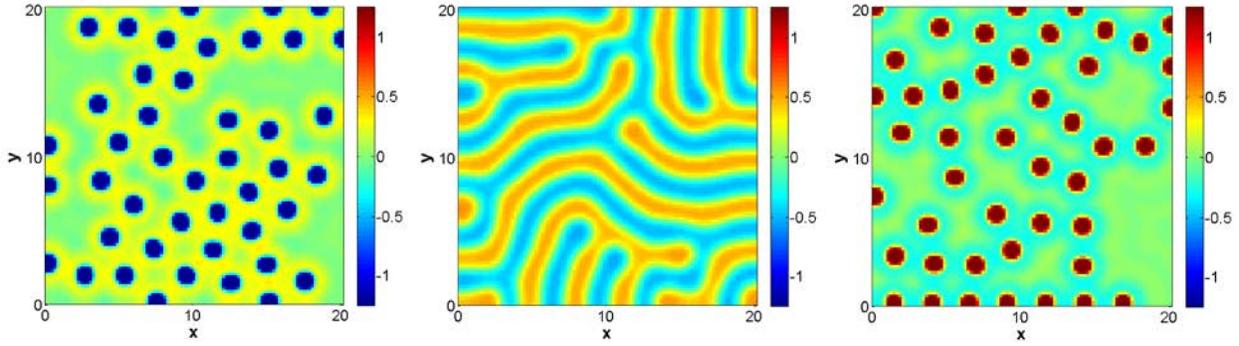

FIG. 5. Sample stationary solution of 2-D Turing model (14) in two spatial dimensions. Panels correspond to (left) inverted spots with q = -2 and r = 4.5, (center) stripes with q = 0 and r = 1, and (right) spots with q = 2 and r = 4.5. All other model/simulation parameters are indicated in the text. Simulations were run until a numerical steady state was reached (maximum difference in any lattice point less than 1e-16, the machine epsilon).

Lastly, we will consider the following Fitzhugh-Nagumo system

$$\frac{\partial u}{\partial t} = D\nabla^2 u - (u-R)(u^2-1) - \rho(v-u)$$

$$\epsilon \frac{\partial v}{\partial t} = \nabla^2 v - (v-u)$$

(15)

where $\epsilon$ and $\rho$ are real valued constants, the diffusion coefficient $D$ is positive and $-1 < R < 1$.

To show the full applicability of our method, for this last study case we will first consider the Fokker-Planck equation, from which we will obtain parameter values that result in one or two maxima. Then, we will incorporate this information into the numerical discretization of the corresponding reaction-diffusion system, corroborating the emergence of the expected patterns.

In cases such as (15), we cannot obtain a closed solution. Correspondingly, we numerically solved the Fokker-Planck equation related to system (15) (cf. Methods section, The Fokker-Planck Associated equation), iterating long enough to consider that a steady state had been reached. We obtained the results shown in FIG. 6. Under these considerations, $\rho$ denotes the probability density distribution, whereas $D$ is the coefficient that characterizes the disturbative strength of the statistical process.

From the numerical solution of the Fokker-Planck equation related to system (15), we obtained one set of parameters that corresponds to each pattern, namely spots or labyrinths. However, it is important to note that these sets are not unique as many combinations of parameters obtained from the Fokker-Planck equation can yield the same qualitative patterns. However, the selected parameters chosen for our simulations were:

(a) Spots: $D = 0.04$; $\epsilon = 1$; $R = 0.04$; $\rho = 0.3$ (inverted spots with identical parameters except $R = -0.04$)

(b) Labyrinths: identical parameters as case (a) above, except $R = 0$.

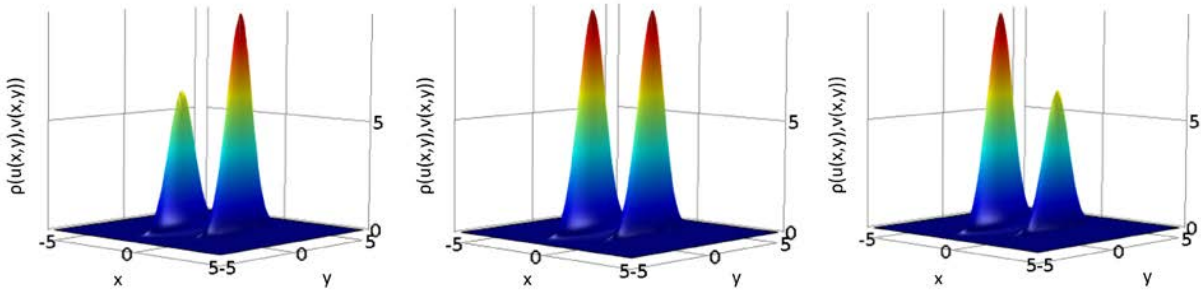

FIG. 6. Associated Fokker-Planck solutions of Fitzhugh-Nagumo model (15), corresponding to two morphogens in two spatial dimensions. Cases correspond to (left) $R = -0.04$; (center) $R = 0$; (right) $R = 0.04$. All other simulation parameters $D = 0.04, \epsilon = 1, \rho = 0.3$ are identical in all panels. The Fokker-Planck equation was solved in Comsol, until steady state (total time T = 1000) with a time step of 0.01, and zero-flux boundary conditions. The initial condition was defined as $u(x, y, t = 0) = e^{-(x^2+y^2)}$ over the domain $x, y \in [-5,5]$.

In order to test these parameters obtained through the Fokker-Planck equation, we discretized the Fitzhugh-Nagumo equations in order to quickly verify if the parameters that corresponded to one or two "wells" in the Fokker-Planck steady state solution would indeed lead to spot- or labyrinth-like patterns. In this case we used a rectangular grid ($[-10,10] \times [-10,10]$ units) with $10^4$ grid points (see Methods section for further details).

As was expected, running numerical simulations for parameters that corresponded to unequal maxima in the Fokker-Planck simulation resulted in spots or inverted spots, whereas solutions of the Fokker-Planck equation that contained equal extrema yielded labyrinthic patterns (FIG. 7).

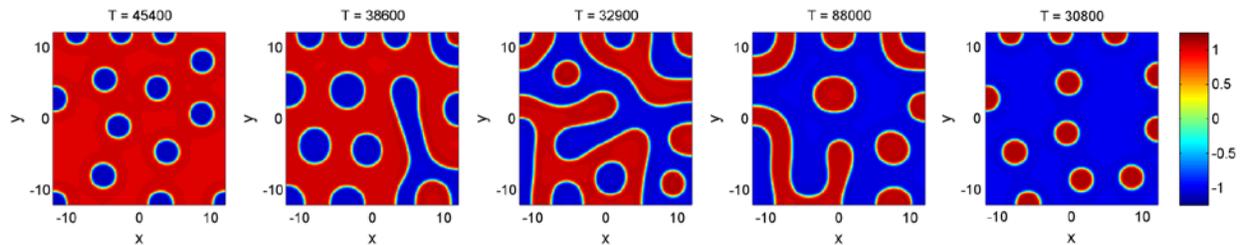

FIG. 7. Sample stationary solution of Fitzhugh-Nagumo system (15), corresponding to two morphogens in two spatial dimensions. Cases correspond to $R = -0.04, -0.02, -0.01, 0, 0.01, 0.02, 0.04$, from left to right. These cases correspond to the transition from inverted spots (R = -0.04, -0.02, -0.01) to labyrinths (R = 0) to spots (R = 0.01, 0.02, 0.04). All other simulation parameters $D = 0.04, \epsilon = 1, \rho = 0.3$ are identical in all panels. Simulations were run until a numerical steady state was reached (maximum difference in any lattice point less than 1e-16, the machine epsilon). Total time until reaching steady states are shown above each case. The color bar included in the right side of the Figure applies to all panels identically.

Moreover, as can be observed in FIG. 8 and FIG. 9, the observed qualitative patterns are reproducible, and our criterion indeed predicts emergent patterns accurately. In FIG. 8, we present representative simulations, where each row corresponds to distinct random initial conditions (used uniformly over simulations within each row), while columns correspond to variations in reaction parameter $R$.

Furthermore, we performed additional simulations to distinguish between effects of diffusion coefficients and reaction rates. FIG. 9 shows simulations using identical random initial condition in all panels, where each row corresponds to a different diffusion coefficient, and columns again correspond to variations in reaction parameter $R$. As one can observe, intermediate and slow diffusion rates yield spatially inhomogeneous emergent patterns, while higher diffusion coefficient values do not.

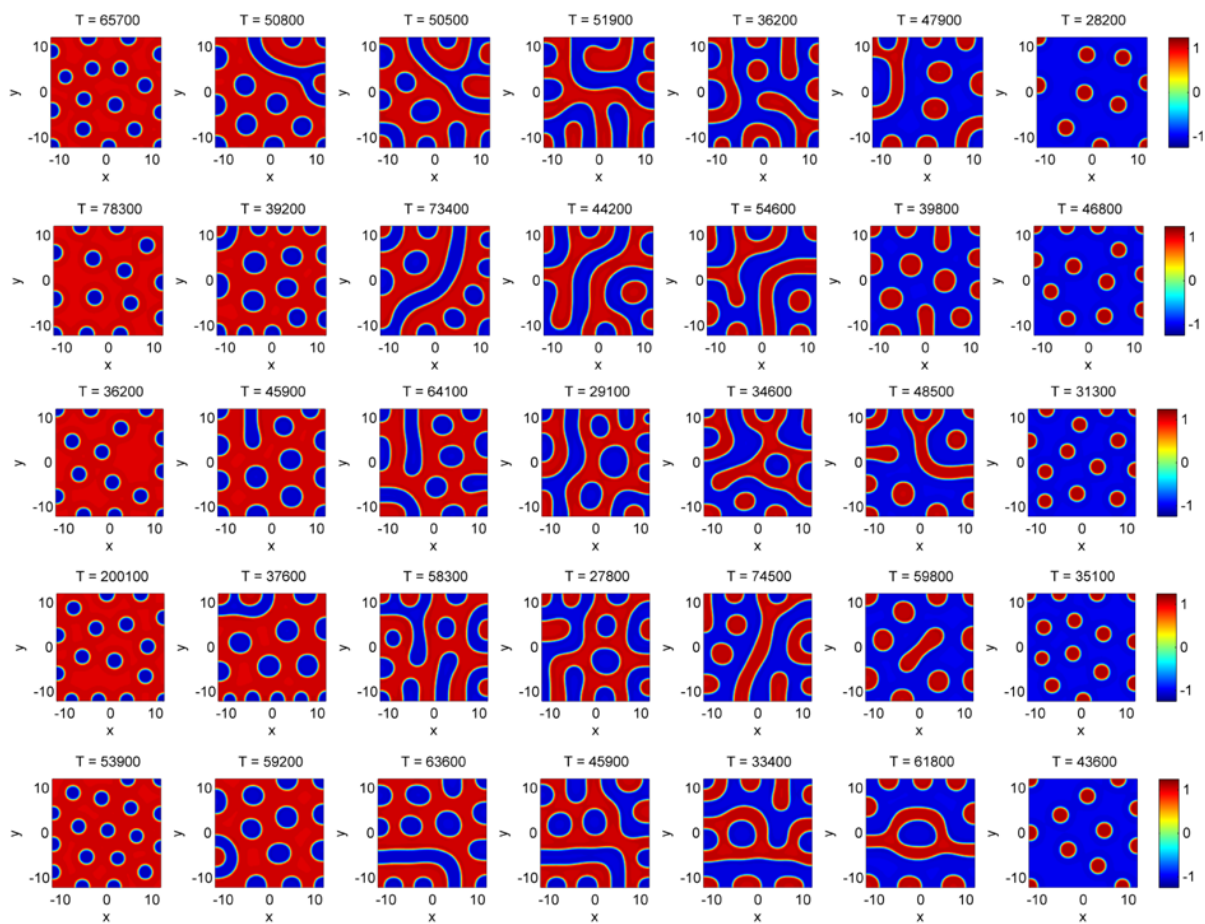

FIG. 8. Sample stationary solutions of Fitzhugh-Nagumo system (15), corresponding to two morphogens in two spatial dimensions. Cases correspond to $R = -0.04, -0.02, -0.01, 0, 0.01, 0.02, 0.04$, from left to right. These cases correspond to the transition from inverted spots (R = -0.04, -0.02, -0.01) to labyrinths (R = 0) to spots (R = 0.01, 0.02, 0.04). All other simulation parameters $D = 0.04, \epsilon = 1, \rho = 0.3$ are identical in all panels. Different rows correspond to uniform random initial conditions used, to ensure comparability between parameter variations. Simulations were run until a numerical steady state was reached (maximum difference in any lattice point less than 1e-16, the machine epsilon). Total time until reaching steady states are shown above each case.

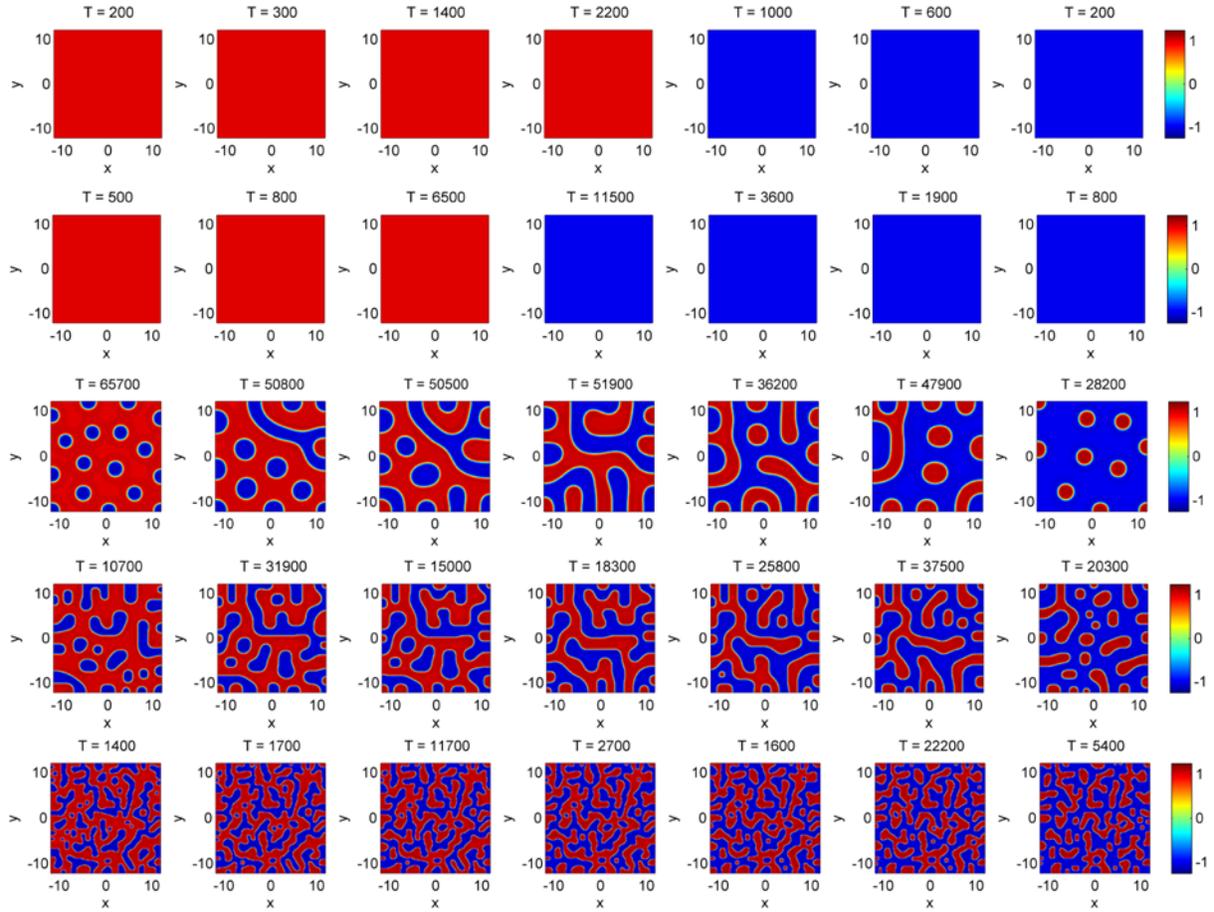

FIG. 9. Sample stationary solutions of Fitzhugh-Nagumo system (15), corresponding to two morphogens in two spatial dimensions. Cases correspond to $R = -0.04, -0.02, -0.01, 0, 0.01, 0.02, 0.04$, from left to right. These cases correspond to the transition from inverted spots (R = -0.04, -0.02, -0.01) to labyrinths (R = 0) to spots (R = 0.01, 0.02, 0.04). Different rows correspond to distinct diffusion coefficients, D = 0.16, D = 0.08, D = 0.04, D = 0.02 and D = 0.01 from top to bottom. Identical random initial conditions were used in all cases, to ensure comparability between parameter variations. Simulations were run until a numerical steady state was reached (maximum difference in any lattice point less than 1e-16, the machine epsilon). Total time until reaching steady states are shown above each case.

Transient spots can slowly reduce their size and/or move through the domain until settling in their stationary position/sizes. As would be expected, systems with parameters corresponding to spot-like solutions can also yield a stationary constant solution if diffusion is fast enough. In contrast, labyrinth-like solutions do not travel throughout the domain. However, if diffusion is fast enough, a stationary constant solution will also be obtained, the value of which will depend on random initial conditions used (data not shown). These observations can in fact be theoretically expected. When the diffusion constant $D$ is high (e.g. system (15) with $D = 0.08\ or\ 0.16$), the diffusion rates of morphogens are not different enough to generate a diffusion-driven instability. Differences between optima of the Fokker-Planck equation in each case are subsequently shown in FIG. 10.

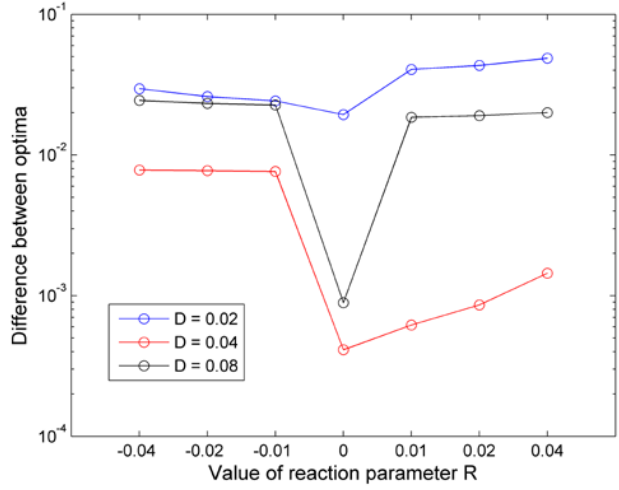

FIG. 10. Absolute differences between optima of Fokker-Planck equations associated to the Fitzhugh-Nagumo model (15). Color-coded curves correspond to distinct diffusion coefficients of morphogen $u$, while points within each curve correspond to distinct values of reaction parameter $R$. In all cases, the Fokker-Planck equation was solved in Comsol, until steady state (total time T = 1000) with a time step of 0.01, and zero-flux boundary conditions. The initial condition was defined as $u(x,y,t=0) = e^{-(x^2+y^2)}$ over the domain $x, y \in [-5,5]$.

### 2.4 Lyapunov function

It is important to notice that when the system has variational structure an analytical statement can be made (interested readers should refer to the calculus of variations section, for essentials on the topic). In such case, we can assume that there is a function $\varphi$, a potential, such that

$$u_t = D_1 \Delta u - \varphi_u(u,v)$$
$$v_t = D_2 \Delta v - \varphi_v(u,v)$$
(16)

The system can be viewed as the Euler-Lagrange equation of the functional

$$\int |\nabla u|^2 + |\nabla v|^2 + \varphi(u,v)\, dx \tag{17}$$

Moreover, this fact guarantees that any region which is invariant for the system

$$\dot{u} = -\varphi_u(u,v)$$
$$\dot{v} = -\varphi_v(u,v)$$
(18)

is also invariant for the reaction-diffusion system in the $L^\infty$ norm (see [25]). In particular, any sublevel set of $\varphi$

$$I^C = \{(u,v) \mid \varphi(u,v) \leq C\} \tag{19}$$

is invariant for the differential equation (DE) system since it is by assumption a Lyapunov function. Notice that in the scalar case

$$\rho = e^{\varphi/\varepsilon} \tag{20}$$

would be a solution to the corresponding Fokker-Planck equation, where $\varepsilon$ is understood as above and, therefore, our criterion reduces to the one already explained. For the system case, if we assume that $\varphi(u,v) = \varphi_1(u) + \varphi_2(v)$, one can directly check that

$$e^{\frac{\varphi_1}{d_1} + \frac{\varphi_2}{d_2}} \tag{21}$$

is the solution of the associated Fokker-Planck equation, where $d_1$ and $d_2$ are given in terms of $D_1$ and $D_2$ above, respectively [30].

Therefore the same observation about the structure of the critical points of $\rho$ and $\varphi$ that we made for the scalar case are valid here, and the criterion is rigorous in this case.

From the physical perspective, the above criterion can be justified in terms of an energy landscape for a two phase system, as discussed in the introduction. Having presented the details, it becomes clear that $-\rho$ plays the role of some effective potential for the system.

## 3 DISCUSSION

We have provided both analytical and numerical evidence supporting the fact that the solution of the Fokker-Planck equation associated with a Turing system can be used in order to determine the type of emergent pattern. The following conjecture seems natural, and describes our selection criteria in a nutshell:

*Assume that the stationary solution of the Fokker-Planck equation (9) has exactly two global maxima of different (equal) height. Then system (1) admits a spot-like (a stripe-like) solution.*

Our presented methodology provides a powerful yet straightforward way to identify parameters yielding specific spatio-temporal emergent patterns. This is particularly important due to the prohibitive cost of computational parameter searches. Moreover, our approach is not limited by the number of equations or spatial dimensions, and general non-linearities can be taken into consideration. Hence, we believe our parameter selection criteria will greatly aid the creation of models of morphogenesis, including applications in developmental biology and chemical patterns.

On the theoretical side, and in terms of future work, it should be noted there are standard techniques that guarantee the existence of invariant regions, and some of these involve the existence of a Lyapunov function. The existence of connecting orbits between two different equilibria or from an equilibrium to itself (homoclinics or heteroclinics respectively, see [31] for details) has been proved using topological techniques, considering information about the nature of the flow along the boundary of the invariant region (see the chapters on the Conly index in [25]). Thus, it seems natural to use the negative of the solution of the Fokker-Planck equation. A rigorous proof of our methodology, in this context, is current work in progress, and will be presented elsewhere.

Aside, a more detailed analytical study is needed in order to make the result applicable in more general situations. For instance, when considering heterogeneous boundary conditions, or when the domain size is not considered to be constant. Another issue worth exploring is the description of patterns as phase transitions. In other words, it seems reasonable to visualize the different patterns that emerge (from spots first, to stripes, and then to inverted spots) as parameters change as some kind of phase transition.

# 4 METHODS

## 4.1 Numerical discretization of reaction-diffusion system

Our code was built in MatLab, using finite differences and backward Euler time stepping. Simulations were run until deviations between sets of time steps were negligible. Namely, until the maximum difference between the current and previous time step of any point is less than $10^{-10}$. In some cases, such maximum difference criterion was stringently reduced to $10^{-16}$, (the epsilon of the machine) to guarantee reaching a numerical stationary solution. The Matlab built-in numerical solution of the Laplacian was used, and the left-hand side (LHS) of the discretization (i.e. the time derivative) was solved with a sparse incomplete Cholesky factorization with a tolerance of $10^{-10}$. For the right-hand side (RHS), i.e. the nonlinearity and coupling factor, $u$ and $v$ were solved using preconditioned conjugate gradients at every time step, with a tolerance error of $10^{-10}$. Also, zero-flux boundary conditions were enforced at every time step. All simulations were run with random initial conditions. Namely $u$ (one morphogen case), or $u$ and $v$ (two morphogens case) were assigned uniformly distributed values between -1 and 1, at each spatial discretization point. Therefore, no pre-pattern was utilized.

## 4.2 The Fokker-Planck associated equation

For a general two equation Turing system, such as (1), we obtain the following Fokker-Planck equation (cf. Reference [29] for details):

$$\frac{\partial \rho}{\partial t} + \frac{\partial}{\partial u}[f(u,v)\rho(t,u,v)] + \frac{\partial}{\partial v}[g(u,v)\rho(t,u,v)] = \frac{1}{2}\left[\frac{\partial^2 2D_1\rho(u,v)}{\partial u^2} + \frac{\partial^2 2D_2\rho(u,v)}{\partial v^2}\right] \quad (B1)$$

where $\int_{-\infty}^{\infty}\int_{-\infty}^{\infty} \rho(u,v)du\,dv = 1$, for which we will obtain the stationary state solution.

## 4.3 Calculus of variations

The calculus of variations is a broad subject dealing with optimization problems in infinite dimensions. Its applications range from the study of geodesics and other geometric problems to control theory in chemistry, economics, engineering, physics, etc. For a modern introduction we refer to P. Pedregal's book ([32]). Here, for the sake of completeness, and for readers not familiar with this result, we present an informal derivation of the Euler-Lagrange equation corresponding to the energy functional

$$E(u) = \int_\Omega |\nabla u|^2 + F(u)\,dx \quad (C1)$$

where $E$ is defined for functions $u(x)$ on $\Omega$, a domain in some $n$-dimensional Euclidian space. For simplicity $\Omega$ is assumed to be regular and bounded, and we will consider smooth functions $u$ which vanish on its boundary.

The Euler-Lagrange equation is a necessary condition for an interior minimum of $E$ to be attained at a certain function $v$. More precisely, if $E$ has an interior local minimum at $v$, then this function satisfies the Euler-Lagrange equation

$$\nabla^2 v - f = 0 \tag{C2}$$

in $\Omega$ with Dirichlet boundary conditions, that is $v = 0$ on the boundary, $\partial\Omega$.

This equation is equivalent to the standard condition of the vanishing of the derivative for functions of one variable at point that constitute interior maxima or minima.

The proof follows directly from the observation that if $v$ is a minimum of

$$E(u) = \int_\Omega |\nabla u|^2 + F(u) \, dx, \tag{C3}$$

then

$$E(v + \varepsilon\,\phi) \geq E(v) \tag{C4}$$

for any fixed $\phi$ that vanishes on the boundary, and $\varepsilon$ sufficiently small. In particular, this implies that as a function of $\varepsilon$ ($\phi$ and $v$ being fixed), and $\varepsilon = 0$,

$$\frac{d}{d\varepsilon} E(v + \varepsilon\varphi) = 0 \,. \tag{C5}$$

Computing the previous derivative after a standard application of Leibnitz rule and integration by parts yields

$$E(v) = \int_\Omega \left(\nabla^2 v - F'(v)\right)\varphi \, dx = 0. \tag{C6}$$

Since the previous equality holds for arbitrary $\varphi$, it follows that the term in parentheses has to vanish identically, which gives the equation used in the paper.

## ACKNOWLEDGMENT

TML was supported by CONACYT (Mexico, Consejo Nacional de Ciencia y Tecnologia, grant number 148724). For PP, this work was partially supported by CONACYT (projects 3703P-E9607 and G25427-E).